\documentstyle[preprint,epsf,aps]{revtex}
\begin{document}
\title{ On the average fidelity criterion for the non-classicality of the continuous variable quantum teleportation}
\author{Wang Xiang-bin\thanks{email: wang$@$qci.jst.go.jp} 
\\
        Imai Quantum Computation and Information project, ERATO, Japan Sci. and Tech. Corp.\\
Daini Hongo White Bldg. 201, 5-28-3, Hongo, Bunkyo, Tokyo 113-0033, Japan}

\maketitle 
\begin{abstract}
We show that the existing argument on the non-classicality of the continuous variable quantum teleportation
(CVQT) experiment by the average fidelity criterion 
is incomplete therefore so far it is still unclear whether the CVQT experiment(Furusawa et al, Science,282, 706(1998)) 
is really non-classical.
\end{abstract}
The main argument on why the continuous variable quantum teleportation experiment\cite{furu} is non-classical
is that it can teleport the coherent states with an average fidelity larger than $\frac{1}{2}$. As it is shown
in Ref.\cite{jmo}, given an ensemble of coherent states $\{|\beta\rangle\}$ which distribute uniformly in the whole complex
plane for $\beta$, if one takes a measurement on each states and obtain an outcome $|\alpha\rangle$,
in principle the average fidelity $|\langle\alpha|\beta\rangle|^2$ never exceeds $\frac{1}{2}$. Therefore,
if certain scheme can teleport the coherent states which are uniformly distributed in the whole complex plane and
the average fidelity between the outcome state and the original state is larger than $\frac{1}{2}$, it must be non-classical. 
Unfortunately, this conclusion is not applicable to the experimental results in Ref.\cite{furu}.
In principle, in any experiment, no one is able to prepare an ensemble of coherent states which distribute uniformly 
in the whole complex plane, because doing so requires an infinite large energy. 
In the experiment\cite{furu}, the ensemble of states used is a uniform distribution over a {\it finite} 
area around the original point. This violates the condition
required in applying average fidelity criterion, which requires the coherent states teleported be uniformly
distributed in the {\it whole} complex plane.
 It has never been shown anywhere
why a uniform distribution over that {\it finite} area can approximate 
the uniform distribution over the {\it whole} complex plane.
Neither has it been shown under which condition one can reasonably take such an approximation or 
how good  the approximation is.
Note here a uniform distribution is quite different from a Gaussian distribution of $\exp(-\lambda |\beta|^2)$ with a
positive parameter $\lambda$. In a Gaussian distribution with $\lambda > 0$, in principle one can always find a
finite area around the original point so that the total probability outside the area is very small therefore the finite
area approximates the whole complex plane very well. However, for a uniform distribution, i.e., $\lambda=0$, 
this is not true.
No matter how large a finite area is chosen, the total probability outside the area is always close to 1 therefore the outside
area can never be ignored without an appropriate additional argument. 

As it has been shown in\cite{jmo}, given a coherent state $|\beta\rangle$ with the prior
probability distribution  $p(\beta)=\frac{\lambda}{\pi}\exp(-\lambda |\beta|^2)$, one may use the
POVM $\{\hat E_\alpha\}$ to maximally estimate the state, with 
\begin{eqnarray}
\hat E_\alpha=\frac{1}{\pi}|\alpha\rangle\langle\alpha|,
\end{eqnarray}
and then conclude  the state to be $|f_\alpha\rangle$ with 
the outcome $\alpha$. In such a strategy, the achievable fidelity is
\begin{eqnarray}
F(\lambda)=
\frac{\lambda}{\pi^2}
\int\langle f_\alpha
\left[\int|\exp\left(-\lambda |\beta|^2-|\alpha-\beta|^2\right)|\beta\rangle\langle\beta|{\rm d}^2\beta 
\right]
|f_\alpha\rangle {\rm d}^2\alpha, 
\end{eqnarray}
where the integrations are all over the whole complex plane. The maximum value  $F_{max}(\lambda)$
is the upper bound of the fidelity by  any classical teleportation.
It has been shown that, in the limit $\lambda\rightarrow 0$, the upper bound for $F(0)$ is 
$\frac{1}{2}$. However, this upper bound is not applicable to the judging of the non-classicality
of  the experiment in Ref.\cite{furu}. In that experiment, the coherent states prepared for teleportation
are uniformly distributed in a finite area $A$ around the original point in the complex plane $\beta$. In such situation,
the upper bound of the classical teleportation fidelity is the maximum value of
\begin{eqnarray}
F'(\lambda)=\frac{\lambda}{\pi^2}
\int\langle f_\alpha
\left[\int_A|\exp\left(-\lambda |\beta|^2-|\alpha-\beta|^2\right)|\beta\rangle\langle\beta|{\rm d}^2\beta \right]
|f_\alpha\rangle 
{\rm d}^2\alpha
\end{eqnarray}
in the limit of $\lambda\rightarrow 0$. Note that here the integration on the complex variable $\beta$ is over a finite
area $A$ instead of the whole complex plane. We can rewrite the above formula in the following  equivalent form
\begin{eqnarray}
F'(\lambda)=F(\lambda)-\frac{\lambda}{\pi^2}
\int\langle f_\alpha
\left[\int_{\bar A}|\exp\left(-\lambda |\beta|^2-|\alpha-\beta|^2\right)|\beta\rangle\langle\beta|{\rm d}^2\beta 
\right]
|f_\alpha\rangle {\rm d}^2\alpha
\end{eqnarray}
and $\bar A$ indicates the area outside $A$ in the complex plane. In the limit of $\lambda\rightarrow 0$, the integration
over $\bar A$ cannot be ignored because no matter how large $|\beta|$ is, the value $|\alpha-\beta|^2$ can be still
rather small. Actually, the whole exponential term never disappears.
Therefore we have no reason to believe the maximum value of $F'_{max}(\lambda\rightarrow0)$ 
is close to the value  $F_{max}(\lambda\rightarrow 0)$.

Although we do not know the exact value of $F'_{max}(\lambda\rightarrow 0)$, qualitatively, 
we know that the value $F'_{max}(\lambda=0)$ must be larger $\frac{1}{2}$, because in such a case
one has more prior information than that in the case that the coherent states are distributed over the whole complex plane.
Consequently it is incorrect to claim that any average fidelity exceeds the bound $\frac{1}{2}$ must have come about through the entanglement
in any real experiment. 
The fidelity by the CVQT experiment in Ref.\cite{furu} is 0.58. So far it is not clear whether
$F'_{max}(\lambda=0)-F_{max}(\lambda=0)$ is larger than 0.8 therefore it is unclear whether the experiment in non-classical.

We are definitely not to claim that there is no way to verify the non-classicality of the 
CVQT experiment\cite{furu}. 
Neither do we claim that to demonstrate the non-classicality one must really prepare an ensemble of coherent states over the whole
complex plane.
To strictly demonstrate the non-classicality of the CVQT experiment, one can use the criterion for the 
Gaussian distribution\cite{jmo} with $\lambda>0$ instead of a uniform distribution of the coherent  states.
In such a Gaussian distribution, we can use a finite area to approximate the whole complex plane very well provided the total
probability outside the area is negligible.
To demonstrate the non-classicality, the existing experimental 
data in Ref.\cite{furu} can be treated in the following way: 
First find a real value $\lambda > 0$ so that the integration for the function $\exp(-\lambda |\beta|^2)$
outside the experimental area is almost 0. Then calculate the average fidelity in the teleportation experiment
weighted by the factor $\frac{\lambda}{\pi}\exp(-\lambda |\beta|^2)$ for each case. Compare the weighted average fidelity with the value 
$\frac{1+\lambda}{2+\lambda}$\cite{jmo}. 
If the former one is indeed larger than the later one, a non-classical teleportation experiment is verified. 

In summary, we have shown that the average fidelity criterion of $F>\frac{1}{2}$\cite{jmo} is not applicable to the CVQT experiment\cite{furu}.
The average fidelity criterion requires the coherent states to be teleported be uniformly distributed in the {\it whole} complex plane while
the real experimental result of the average fidelity is  obtained 
with a uniform distribution  over a {\it finite} area of the complex plane.
To know whether the CVQT experiment in Ref.\cite{furu} is
really non-classical, a further study is needed.

{\bf Acknowledgement:} 
I thank Prof Imai H for support. I thank Dr Fan Heng, Dr. Matsumoto K and Dr. 
for useful discussions. 

\end{document}